\documentclass[english,graphics,floatfix,superscriptaddress]{revtex4-1}
\usepackage{graphicx}
\usepackage{subfig}
\usepackage{amssymb,amsmath,amsfonts,mathrsfs,eufrak,fancyhdr}
\usepackage{hyperref}
\usepackage{color}
\usepackage{makeidx,endnotes}
\usepackage{showidx}
\usepackage{theorem,eepic}
\usepackage{epsf,epsfig}
\usepackage{epstopdf}

\def\<{\langle}
\def\>{\rangle}

\def\set#1{{\sf #1}}

\def\map#1{{\mathcal{#1}}}

\begin{document}
\title{Universality in random quantum networks}
\author{Jaroslav Novotn\'y}
\affiliation{Department of Physics, Czech Technical University in Prague, Faculty of Nuclear Sciences and Physical Engineering, B\v rehov\'a 7, 115 19 Praha 1 - Star\'e M\v{e}sto, Czech Republic}
\author{Gernot Alber}
\affiliation{Institut f\"ur Angewandte Physik, Technische Universit\"at Darmstadt, D-64289 Darmstadt, Germany}
\author{Igor Jex}
\affiliation{Department of Physics, Czech Technical University in Prague, Faculty of Nuclear Sciences and Physical Engineering, B\v rehov\'a 7, 115 19 Praha 1 - Star\'e M\v{e}sto, Czech Republic}

\date{\today}
\begin{abstract}
Networks constitute efficient tools for assessing universal features of complex systems. In physical contexts, classical as well as quantum, networks are used to describe a wide range of phenomena, such as phase transitions, intricate aspects of many-body quantum systems or even characteristic features of a future quantum internet. Random quantum networks and their associated directed graphs are employed for capturing statistically dominant features of complex
quantum systems. Here, we develop an efficient iterative method capable of evaluating the probability of a graph being strongly connected. It is proven that random directed graphs with constant edge-establishing probability are typically strongly connected, i.e. any ordered pair of vertices is connected by a directed path. This typical topological property of directed random graphs is exploited to demonstrate universal features of the asymptotic evolution of large random qubit networks. These results are independent of our knowledge of the details of the network topology. These findings suggest that also other highly complex networks, such as a future quantum internet, may exhibit similar universal properties.
\end{abstract}
\maketitle

\section{Introduction}
Networks are important models of dynamical systems both in quantum and classical physics with applications ranging from many-body physics and communication technology
to social sciences and economics \cite{Barabasi2002,Cohen2010,Newman2001}. The rapid advancement of quantum information technology \cite{Bruss2007} has stimulated considerable interest in dynamical properties of quantum networks formed by elementary systems, such as qubits, in view of their privileged role in quantum communication and quantum computation \cite{Chuang2000}.
In particular, qubit based quantum networks may even model
characteristic features of a future quantum internet and of the resulting intricate possibilities of generating and distributing entanglement among its nodes. In a typical model of a quantum internet \cite{Preskill1999,Kimble2008,Hughes2013,Bussieres2014} distinguishable quantum systems, such as material qubits realized by trapped ions or atoms, are coupled by photon exchange through optical fibers. These nodes may influence each other by directed entanglement operations which typically
involve a control and a target system and are thus representable by directed links.
At any time the 'users' of such a quantum internet can establish links between these nodes. From the point of view of the physical systems constituting the nodes the establishment of 'user'-controlled connections and subsequently executed entanglement operations
may appear random so that a quantum internet may be viewed as
a paradigm of a random quantum communication network.
Other important examples of quantum networks include systems and processes encountered in quantum statistical physics
which involve approaches to equilibria or stationary states in macroscopic systems.

Recently, intriguing relations between properties of quantum communication networks with prescribed lattice topologies and statistical physics have been reported. Based on classical percolation concepts \cite{Broadbent1957,Grimmett1999,Bollobas2006} it has been shown \cite{Acin2007} that these quantum networks may exhibit an entanglement percolation phase transition \cite{Sachdev1999}. Accordingly, it requires a minimal critical amount of local entanglement in order to establish an almost perfect distant quantum channel through such a network. This early result on quantum networks and its connection to statistical physics \cite{Acin2007} stimulates natural questions concerning the existence and character of topological and dynamical properties of other quantum networks.

Here we address two such questions elucidating characteristic properties of quantum networks with randomly established links and
without any particular prescribed lattice topology \cite{Acin2007}.
The first question addresses the random establishment of directed links which may capture the asymmetric role played by
control and target nodes in entanglement creating operations, for example:
What is the typical topological structure of random networks with
directed links which can be established with equal probability between any two nodes?
Answering this question constitutes a first step
towards an understanding of generic features of entanglement generation and distribution within such random quantum networks. For this purpose we develop an iterative method for determining
the probability that a random directed graph describing such a random network is strongly connected for any edge-establishing probability $p$. As a main result it is demonstrated that almost certainly a sufficiently large random network with directed
links is strongly connected, i.e. there is always at least one closed
directed path through any pair of nodes. Our second question addresses
consequences of this typical topological property on
dynamical features: Does this typical topological
property imply universal asymptotic dynamical features on randomly generated quantum networks? Based on previous results
\cite{Novotny2010,Novotny2011} it is demonstrated that such universal properties arise. As an example the asymptotic dynamics of iteratively and randomly applied entanglement operations is investigated and it is shown that it is independent of most of the details of the underlying randomly generated qubit network. In particular the individual connectivity of the networks nodes is not the crucial feature for the asymptotics.

This
 paper is organized as follows. In Sec. \ref{chapter 2}
the iterative procedure is presented for evaluating the probability that a random directed graph with constant edge-generating probability $p$ is strongly connected. Based on this result in chapter \ref{chapter_3}
the asymptotic dynamics of qubit networks with random unitary evolution is studied. In particular, two different scenarios are investigated, namely a static random network and a dynamically changing random quantum network.

\section{Strong connectivity of large random directed graphs}
\label{chapter 2}
In this section we
investigate the connectivity of random directed graphs which describe random
networks with a constant edge-generating probability $p$.
As a main result the probabilistic dominance of strongly connected directed graphs
is proven in the limit of large networks.

\begin{figure}
\includegraphics[width=0.4\textwidth]{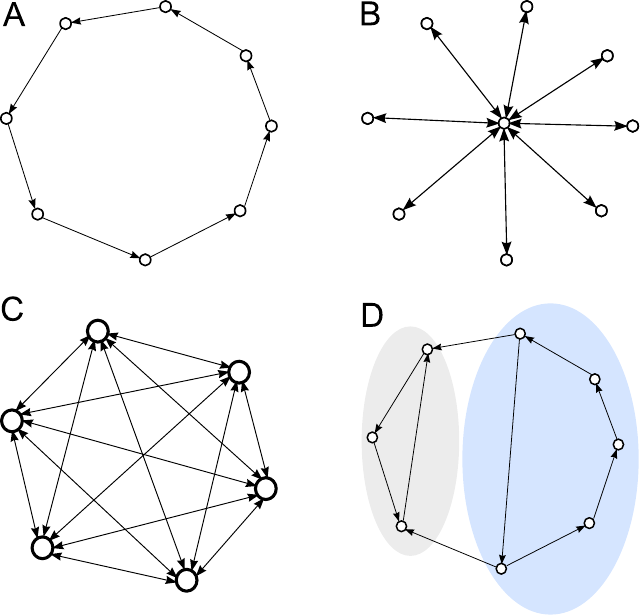}
\caption{Examples of directed graphs: {\bf a} - a
strongly connected cyclic graph, {\bf b} - a
strongly connected star graph, {\bf c} - a
strongly connected complete graph $\map C$, {\bf d} - a
non-strongly connected graph with two strongly connected components
(grey and blue areas).
} \label{fig1}
\end{figure}

The set of all networks with $n$ nodes is characterized
by the set of their associated
directed graphs $\set G(n)$. A graph $g(\set V, \set E) \in \set G(n)$ is defined by its set of vertices $\set V$ representing
the network's
 nodes and by the set of directed edges $\set E$ representing
the network's
directed links.
A complete graph $\map C$ contains all its $n(n-1)$ possible directed edges. We focus on directed graphs in order to be able to take into account that any pair of nodes of the network can be linked in an asymmetric way as in the case of an entanglement operation with a control and a target system. A path in a graph is an ordered set of adjacent edges. A graph is strongly connected if there is at least one closed path through any pair of vertices (compare with Fig. \ref{fig1}). Random graphs are a convenient tool for exploring statistical properties of complex networks. They are defined by a probability distribution on the set $\set G(n)$. In the following we concentrate on random graphs in which all directed $n(n-1)$ edges are treated identically,
i.e. the links are selected randomly and statistically independently with probability $p>0$. Thus $\set G(n,p)$ is the set of directed random
graphs $\set G(n)$ of Erd\"os-R\'enyi type \cite{Erdos1959,Bollobas2001} equipped with the corresponding binomial probability distribution.

A way of classifying topologically the set of random graphs $\set G(n,p)$ is to determine the probability $P_C(n,p)$ of a random graph $g \in \set G(n,p)$ to be strongly connected. This is a challenging graph theoretical problem already for the simpler case of undirected graphs. It has not yet been solved completely \cite{Harary1973,Pittel2012}. In order to tackle this problem we have developed an iterative procedure for evaluating this topological graph theoretical quantity $P_C(n,p)$. Details of this
recurrence procedure are described in the subsequent section.

\subsection{Probability of strong connectedness $P_C(n,p)$}
\label{chapter_2a}
Let us consider directed random graphs with $n$ vertices whose probability of establishing an arbitrary link between any pair of these vertices is given by $0 < p < 1$. The probability that such a graph is not strongly connected is denoted by $P_D(n,p) = 1 - P_C(n,p)$. Any of these directed graphs
can be uniquely decomposed into maximal strongly connected subgraphs, so-called strongly connected dicomponents (SCD). Unlike in undirected graphs these dicomponents may be interconnected provided that connections between different SCDs do not form a directed loop. Note that while each strongly connected graph consists of a single SCD only any non strongly connected or disconnected graph
with $n$ vertices consists of at least two SCDs. Each of them has strictly less than $n$ vertices.

Based on these characteristic properties the whole set of directed disconnected graphs with $n$ vertices and its associated probability can be constructed
by a recursive algorithm described in the following. For this purpose let us choose a partition $\{n\}_k$ of the number of vertices $n$ with length $k \geq 2$. This is a distribution of $n$ vertices into $k$ components described by a decreasing sequence of natural numbers, i.e. $\{n\}_k \equiv \{n_1, n_2, \ldots, n_k\}$ summing up to $n$. Thereby each partition describes one possible decomposition of $n$ vertices into $k$ SCDs.
As different links are selected in a statistically independent way the probability that each of  these $k$ SCDs is strongly connected is given
by the product of the probabilities $P_C(n_i,p); i=1,...,k$. In general, in a disconnected graph these $k$ SCDs may
be interconnected by directed edges which, however, must not form a closed directed loop. Let us denote the probability that these edges do not form a closed directed loop by $P_A(\{n\}_k,p)$. Taking into account all possible distributions of labeled vertices into at least two
SCDs we can determine the probability $P_D(n,p)$ of a directed graph with $n$ vertices being disconnected, i.e.
\begin{eqnarray}
\label{recurrent_probability_scg}
P_D(n,p)&=&\sum_{\{n\}_k \in \map K} K(\{n\}_k) \prod_i P_C(n_i,p) \times P_A(\{n\}_k,p).
\nonumber\\
\end{eqnarray}
The summation includes all possible partitions $\map K$ of $n$ vertices with
a minimum length of two. The number $K(\{n\}_k)$ of possible decompositions of $n$ labeled vertices  within the same partition $\{n\}_k$
can be calculated in a straightforward way.
However, the determination of the probability $P_A(\{n\}_k,p)$ is a formidable combinatorial problem which can also be achieved by
a recursive procedure described in the following.

This recursive procedure starts from the observation
that the absence of any directed loop between different SCDs
implies the existence of at least one SCD which is not adjacent to any outgoing edge connecting it with any other SCD, i.e.  it is not adjacent to any interconnecting edge between different SCDs whose tail lies in this SCD.
Separating this SCD from the rest, the residual SCDs also cannot be interconnected by a directed loop. In principle each SCD of the original set of SCDs could be non-adjacent to any outgoing interconnecting edge into the residual SCDs. Correspondingly, an upper bound on the probability $P_A(\{n\}_k,p))$
is given by  the sum of probabilities associated with all these (not necessarily exclusive) possibilities, i.e.
\begin{eqnarray}
\label{expression_acyclic_graphs}
 P_A\left(\{n\}_k,p\right) &\leq & \sum_{\{m\}_1 \subset {\{n\}_k}} (1-p)^{m(n-m)} P_A\left(\{n\}_k / \{m\}_1,p\right).\nonumber\\
\end{eqnarray}
Thereby, a  SCD with $m$ vertices and without any outgoing edge is represented by
a partition $\{m\}_1$ of unit length and the residual SCDs are represented by the partition
$\{n\}_k /\{m\}_1$ of length $k-1$. The probability of such a graph is given by the product of
the probability $P_A\left(\{n\}_k/\{m\}_1,p\right)$ that the residual SCDs are not strongly connected with
the probability $(1-p)^{m(n-m)}$ that the chosen SCD with $m$ vertices does not have any outgoing edge.
The summation in Eq.(\ref{expression_acyclic_graphs}) includes all $k$ SCDs of the partition $\{n\}_k$.
Thus, it takes into account situations with two or more SCDs, which do not have any outgoing edges, more than once. In order to obtain not only an upper bound but a precise expression for the probability
$P_A(\{n\}_k,p)$ we have to compensate for this overcounting. This can be achieved by subtracting the probabilities describing situations with two SCDs without any outgoing edges, adding the probabilities of situations with three SCDs without any outgoing edges and so on. This way we eventually obtain the recurrence relation
\begin{widetext}
\begin{eqnarray}
\label{probability_acyclic_graphs}
P_A\left(\{n\}_k,p\right) =  \sum_{l < k, \{m\}_l \subset {\{n\}_k}} (-1)^{l-1}(1-p)^{m(n-m)+\sum_{i\neq j}m_i m_j} P_A\left(\{s\}_{k-l},p\right)
\end{eqnarray}
\end{widetext}
for the probability that the SCDs described by the partition $\{n\}_k$
do not contain any closed directed loops connecting them.
The partition $\{s\}_{k-l} \equiv \{n\}_k / \{m\}_l$ refers to the distribution of vertices of the residual $k-l$ SCDs after $l$ SCDs without outgoing links to these $k-l$ SCDs have been separated off.  The sum covers all possible partitions of legth $l$ described  by the partitions
$\{m\}_l$ and chosen from the original partition $\{n\}_k$. From the recurrence relations (\ref{recurrent_probability_scg}) and (\ref{probability_acyclic_graphs}) the probability of a random graph being disconnected and its complementary probability $P_C(n,p)$ can be determined recursively.

\begin{table}[h!]
\begin{tabular}{|l|c|c|c|c|c|c|}
\hline
\bfseries $n$ & 2 & 3 & 4 & 5& 6& 7\\
\hline
\bfseries $P_C(n,1/2)$ & 0.25& 0.3438 & 0.4331 & 0.5742 & 0.7049 & 0.8114\\
\hline
\end{tabular}
\caption{Probability $P_C(n,p=1/2)$ of a graph being strongly connected.}
\label{table_number_scg}
\end{table}

\begin{figure}
\includegraphics[width=0.4\textwidth]{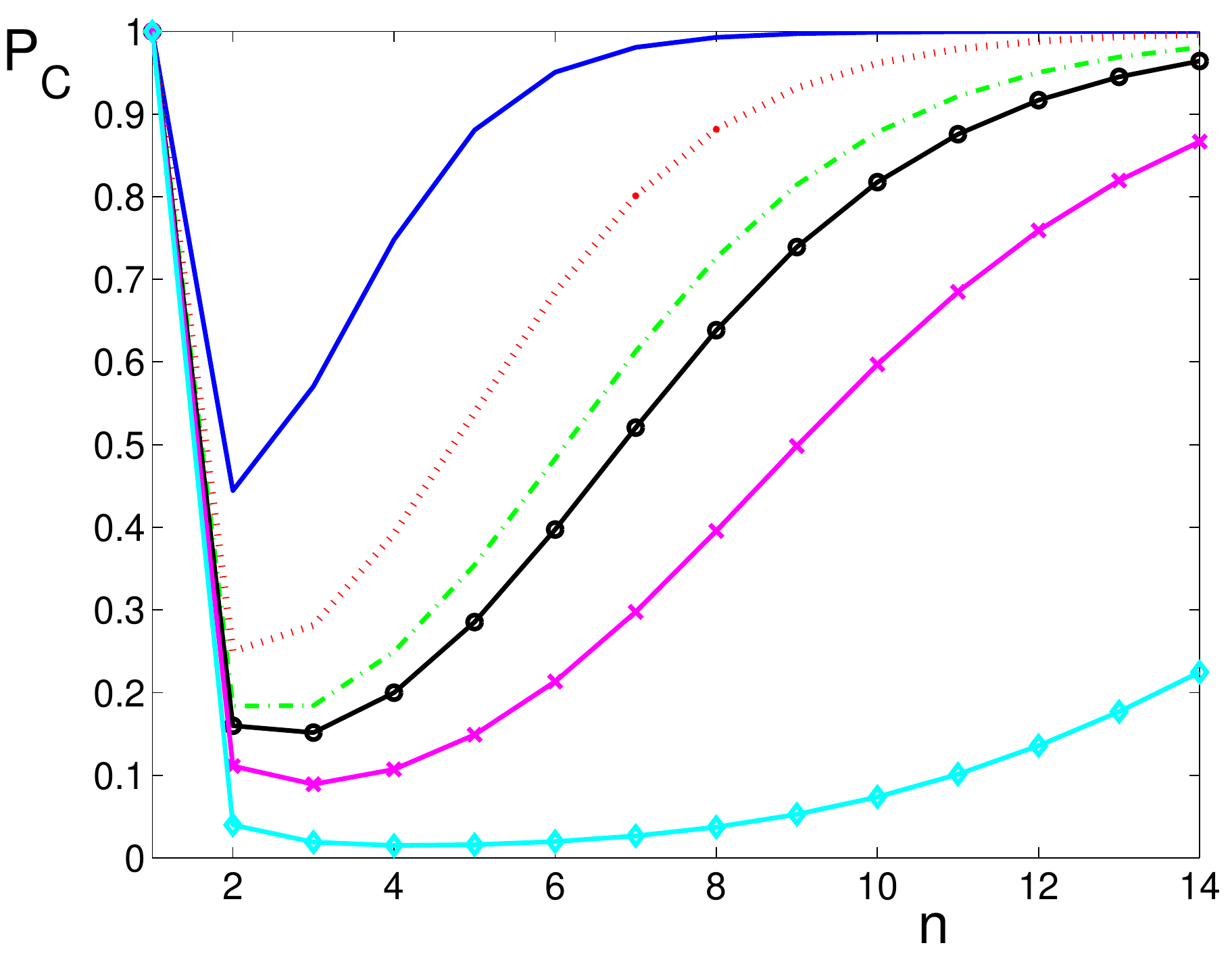}
\caption{Probability $P_C\equiv P_C(n,p)$ for a random directed  graph from the set $\set G(\set V,p)$ to be strongly connected:
$p=2/3$ (full, blue), $p=1/2$ (dotted, red), $p=3/7$ (dash-dotted, green), $p=2/5$ (circle, black), $p=1/3$ (cross, magenta), $p=1/5$ (diamond, cyan).} \label{fig2}
\end{figure}

Employing this recurrence procedure
 we can calculate the probability $P_C(n,p)$.
In Table \ref{table_number_scg} values of $P_C(n,p)$ are shown for $p=1/2$ and for small values of $n$. These results indicate that for small numbers of vertices $n$ it is more likely for a random graph to be non-strongly connected. However, already for $n > 5$ it is more likely that a random graph is strongly connected. In Fig.\ref{fig2} the probability $P_C(n,p)$ is depicted
for various numbers of vertices $n$ and probabilities $p$ of randomly establishing edges. It is apparent that with increasing numbers of vertices $n$ the probability $P_C(n,p)$ approaches unity for any probability
$p$. Fig.\ref{fig2} suggests a general topological result: In the limit of sufficiently large networks a randomly selected directed graph  $g \in \set G(\set n,p)$ is almost certainly strongly connected.

As we show in the following part this main result can be further strengthened quantitatively by
a rigorous lower bound of $P_C(n,p)$ which scales exponentially with the number of vertices $n$.

\subsection{Asymptotic lower bound on $P_C(n,p)$}
\label{chapter_2b}
For large numbers of vertices $n$ the analysis of the behavior of the probability $P_C(n,p)$ based on the recursive relations
(\ref{recurrent_probability_scg}) and (\ref{probability_acyclic_graphs})
constitutes a challenging task \cite{Alimonti,Gimenez}. In the following
we derive the asymptotic lower bound of $P_C(n,p)$
\begin{equation}
P_C(n,p) > 1 - (n-1)^2 (1-p^2)^{(n-1)}
\label{lowerbound}
\end{equation}
which is valid for large numbers of vertices $n$.
It demonstrates that for any probability $0<p<1$ and sufficiently large numbers of vertices $n$ the probability $P_C(n,p)$ always approaches unity exponentially.

An undirected random graph with $n$ vertices, whose probability of establishing an edge between any pair of its vertices is given by $0<p^2<1$. It can be viewed as a special case of a directed graph. In this directed graph the same pairs of vertices are linked in both directions and the probability for establishing a directed link between any two vertices is given by $0<p<1$.
An undirected graph is called connected if any pair of vertices is linked by a path. Apparently connectedness of an undirected graph entails strong connectedness of the associated directed graph and thus yields a lower bound for  $P_C(n,p)$, i.e.
\begin{equation}
\label{rel_oriented_vs_nonoriented}
P_C(n,p) > P_C^{un}(n,p^2).
\end{equation}
Thereby $P_C^{un}(n,p^2)$ denotes the probability that a random graph with $n$ vertices and with probability of edge formation equal to $p^2$ is connected.

Establishing a lower bound on $P_C^{un}(n,p)$ is significantly easier
than evaluating $P_C(n,p)$ because connected components of an undirected graph are not mutually interconnected. Hence, the probability that
a chosen and fixed vertex lies in a (connected) component
with  $k$ vertices is equal to
\begin{equation}
{{n-1}\choose{k-1}}P_C^{un}(k,p)(1-p)^{k(n-k)}.
\end{equation}
The normalization of this probability distribution yields the relation
\begin{equation}
P_C^{un}(n,p)=1- \sum_{k=1}^{n-1}
{{n-1}\choose{k-1}}P_C^{un}(k,p)(1-p)^{k(n-k)}.
\end{equation}
As the probability $P_C^{un}(k,p)$ that the selected $k$ vertices are connected is upper bounded by unity we obtain the lower bound
\begin{equation}
P_C^{un}(n,p) \geq 1- \sum_{k=1}^{n-1}
{{n-1}\choose{k-1}}(1-p)^{k(n-k)} = 1 - \sum_{k=1}^{n-1} a_k .
\end{equation}
The ratio of two consecutive terms $a_k$ and $a_{k+1}$ of this sum is given by
\begin{equation}
\frac{a_{k+1}}{a_k}= \frac{n-k}{k}(1-p)^{n-(2k+1)}.
\end{equation}
It reveals that for sufficiently large values of $n$ the largest element of
this sum, i.e. $a_{n-1} = (n-1)(1-p)^{(n-1)}$, majorizes all other terms.
Thus, for a sufficiently large number of vertices $n$ the probability $P_C^{un}(n,p)$ fulfills the inequality
\begin{equation}
P_C^{un}(n,p) > 1 - (n-1)^2(1-p)^{n-1}.
\end{equation}
Finally, in view of relation (\ref{rel_oriented_vs_nonoriented}) we
obtain a lower bound for the probability $P_C(n,p)$ in the form
of Eq. (\ref{lowerbound}).

This lower bound proves that for any probability $p$ the quantity $P_C(n,p)$ approaches the asymptotic value of unity exponentially fast with increasing number of vertices $n$. From Fig.\ref{fig2} it is also apparent that for each value of $p$ the probability $P_C(n,p)$ attains a minimum as a function of the number of vertices $n$. For smaller values of $p$, the probability distribution becomes shallower in the neighborhood of the minimum. Contrary to the 'macroscopic' regime of large networks, i.e. $n \gg 1$, in the 'mesoscopic'  regime of network sizes close to this minimum the probabilistic dominance of strongly connected directed graphs is lost. It is an interesting issue beyond our scope to explore possible general topological features of random directed  graphs also in this 'mesoscopic' regime.

\section{Universal asymptotic dynamics of strongly connected qubit networks}
\label{chapter_3}
\begin{figure*}[t]
\includegraphics[width=0.9\textwidth]{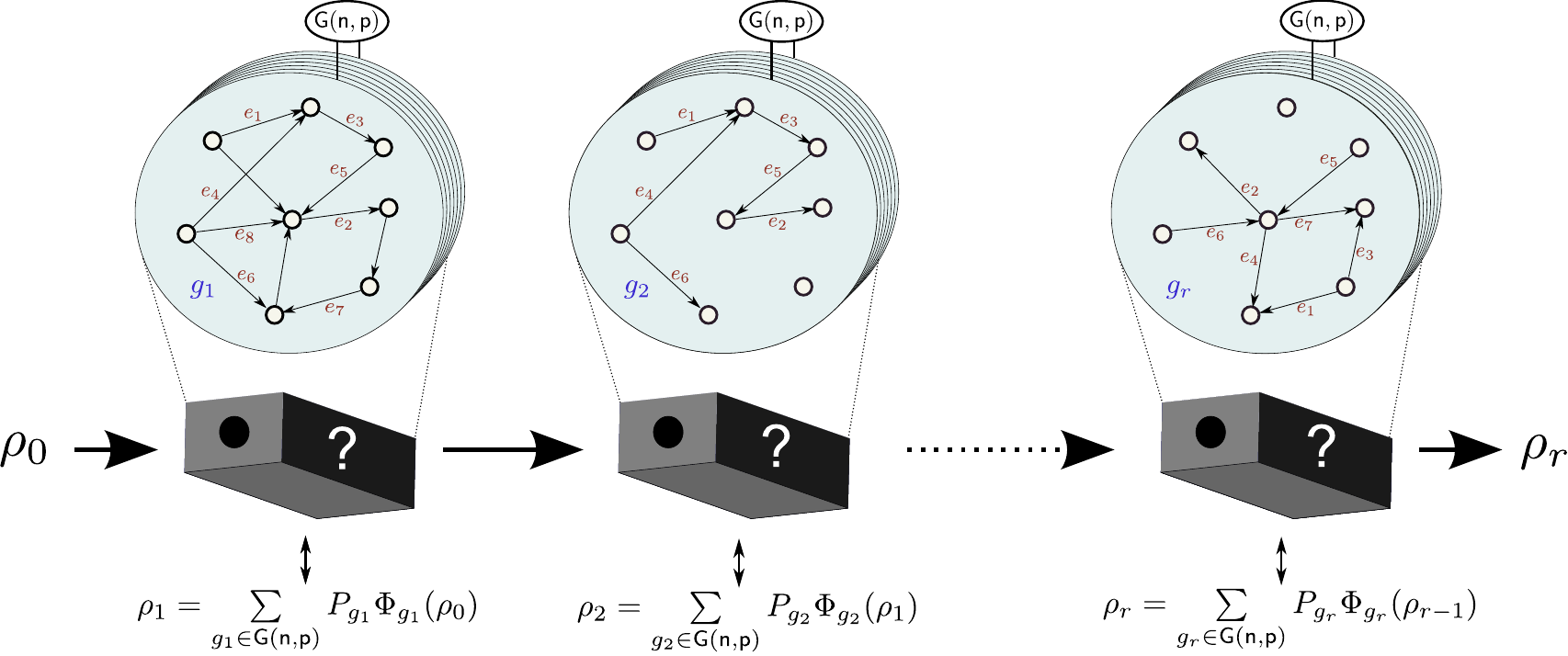}
\caption{Iterative stochastic evolution of a dynamic random quantum network:
In each step of the iteration procedure a
black box generates the random unitary evolution $\Phi_g$
for a particular graph $g \in \set G(n,p)$.
Ignorance of $g$  is described by the quantum operation $\Phi$, i.e. by averaging $\Phi_g$ over the ensemble of random graphs $\set G(n,p)$. A sequence of $r$ black boxes transforms the initially prepared state $\rho_0$ into the state $\rho_r$ iteratively.}
\label{fig3}
\end{figure*}
The topological result of Sec. \ref{chapter 2}
has surprising dynamical consequences.
Let us consider a network characterized by a graph $g\in \set G(n,p)$ whose nodes are formed by distinguishable qubits and whose
links represent all possible pairwise couplings between these qubits. Typical quantum information protocols rely on oriented
sender(S)-receiver(R)
operations involving frequent entanglement generation. Within a future quantum internet with many 'users' such oriented sender-receiver operations along links $l \in g$ are expected to be applied 'unsupervised', i.e. randomly
and independently with probability $q_l>0$. A natural choice for such an oriented sender-receiver operation is the unitary controlled-not
(CNOT) operation
$U_{l=SR}=|0\>_S\<0| \otimes I_R + |1\>_S\<1| \otimes \left(|0\>_R\<1|+ |1\>_R\<0|\right)$ with computational basis $|0\>,|1\>$ and with
the identity operator $I_R$ of the receiver.
The randomness of the applied CNOT operations is captured by the random unitary quantum operation \cite{Kraus1983}
\begin{equation}
\Phi_g(\rho) = \sum_{l\in g} q_l U_l \rho U_l^{\dagger}.
\end{equation}
The density operator of the initially prepared $n$-qubit quantum state is denoted by $\rho$.

Recently, it has been shown that if the chosen graph $g$ is strongly connected iterations of this quantum operation converge asymptotically
towards the state \cite{Novotny2010,Novotny2011}
\begin{equation}
\sigma = \lim_{r \to \infty} \Phi_g^{r}(\rho) =
\map P_g\rho \map P_g + \frac{1 - \eta}{2^n - 2}(I - \map P_g).
\end{equation}
Thereby, $\map P_g$ is the projection operator onto the subspace spanned by the CNOT-invariant pure $n$-qubit states $|0\>^{\otimes n}$ and $(|0\>+|1\>)^{\otimes n}$. The probability with which the initially prepared quantum state $\rho$ is in this subspace is denoted by
$\eta$. The stationary quantum state $\sigma$ is independent of the probability distribution $\left\{q_l>0, l\in g\right\}$. Furthermore, all strongly connected $n$-qubit networks including the completely connected one
converge towards this particular quantum state. Thus, it is the topological property of strong connectedness which determines this stationary quantum state. All other structural properties of the established connections within the qubit network are irrelevant for this universal dynamical property. In a stochastic simulation an individual realization of $r$ iterations of the quantum operation $\Phi_g$ describes a random process in which in each of the $r$ steps a new CNOT operation is selected randomly along one of the links $l\in g$. Averaging these individual realizations at each step over the probability distribution $\left\{q_l>0, l\in g\right\}$ yields the asymptotic quantum state $\sigma$.

Let us now consider a different scenario in which the graph $g\in \set G(n,p)$
can change randomly and independently between iterations. The resulting stochastic evolution describes a dynamic random network
in which one is ignorant of the actual graph (see Fig. \ref{fig3}).
This corresponds to averaging the quantum operation $\Phi_g$ over all randomly generated graphs $g\in \set G(\set n,p)$ with the binomial probability distribution $P_g$ generated by $p$.  It is described by the quantum operation
\begin{equation}
\Phi (\rho) = \sum_{g\in G(V,p)} P_g \Phi_g (\rho).
\end{equation}
Iterating this quantum operation drives the $n$-qubit network also towards the state $\sigma$. This universal behavior is due to the fact that this quantum operation $\Phi$ may be viewed as a random unitary operation $\Phi_{\map C}$ applied along the links of the complete directed graph $\map C$ with $n$ vertices.

\begin{center}
\begin{figure*}
\includegraphics[width=1.0\textwidth]{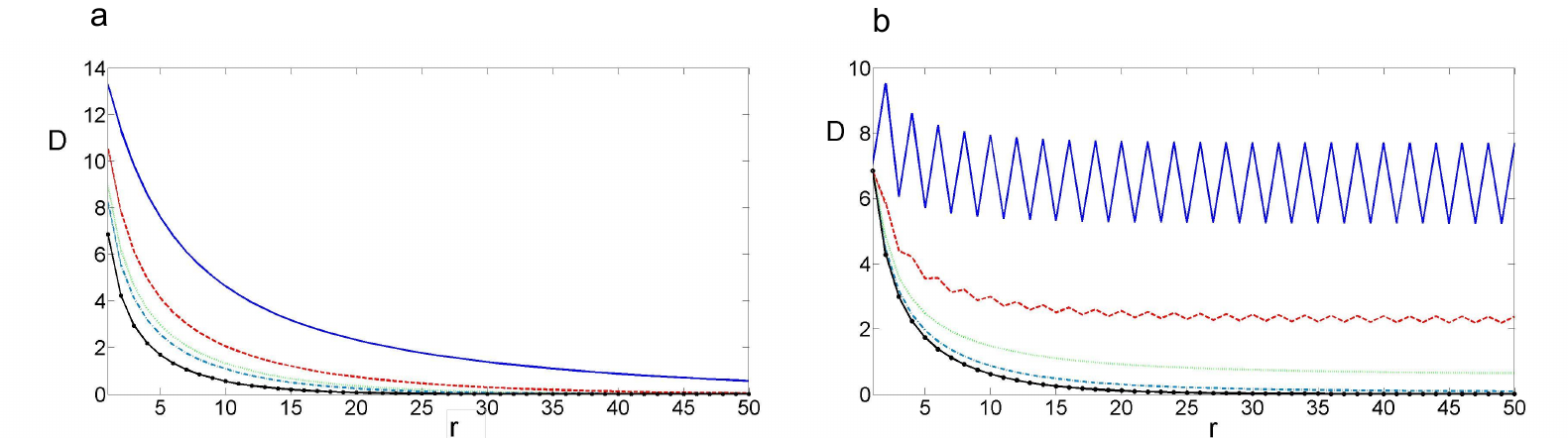}
\caption{Hilbert-Schmidt distance $D$ between the $r$-th iterate $\chi^{(r)}$ of a quantum operation and the asymptotic quantum operation $\Phi_g^{(\infty)}$ of a strongly connected graph $g$ with $n=4$ vertices: {\bf a:} $\chi^{(r)} = \Phi^{(r)}$ (dynamic random networks), {\bf b:} $\chi^{(r)} = \Psi^{(r)}$ (static random networks); the probabilities for generating a directed random link are: $p=0.2$ (blue, solid), $p=0.4$ (red, dash), $p=0.6$ (green, dot), $p=0.8$ (purple, dash-dot), $p=0.95$ (black, solid-diamond).}
\label{fig45}
\end{figure*}
\end{center}

Another stochastic evolution arises if one is ignorant of the graph $g \in \set G(n,p)$ but this graph $g$ remains unchanged during iterations.
This case describes a static random network. In contrast to the previously discussed cases the $r$-th iterate of the corresponding quantum operation $\Psi$ is given by
\begin{equation}
\Psi^{(r)}(\rho) = \sum_{g \in \set G(n,p)} \Phi_g^r (\rho).
\end{equation}
According to the recent result of Ref. \cite{Novotny2011} the quantum state $\Phi_g^{r}(\rho)$ converges towards $\sigma$ if and only if the chosen graph $g \in \set G(n,p)$ is strongly connected. However, due to the probabilistic dominance of strongly connected graphs within the set $\set G(n,p)$ the randomly selected graph $g$ will almost certainly be strongly connected provided the number of vertices $n$ is sufficiently large. Thus, the universal dynamical property of randomly applied CNOT operations \cite{Novotny2011} combined with the probabilistic dominance of strong connectedness in randomly generated large
networks yields the following  general result: On a sufficiently large randomly generated $n$-qubit network the iterated asymptotic dynamics of randomly applied CNOT operations almost certainly converges towards the $n$-qubit state
$\sigma$ of Eq.(2). This state is uniquely determined by the $n$ qubits involved and is independent of all further details of the  network.

Characteristic features demonstrating these universal convergence
properties are illustrated in Figs. \ref{fig45}a,b.
Thereby, the deviations
of the $r$-th iterate of the quantum operations
$\Phi^{(r)}$  and $\Psi^{(r)}$
from the asymptotic dynamics $\Phi_g^{(\infty)}$ of a strongly connected graph $g$
are investigated. These quantum operations describe
the iterated dynamics of a dynamic
and of a static random network.
As a state independent measure of these deviations the Hilbert-Schmidt distances between these quantum operations are depicted for graphs with $n=4$ vertices and with different probabilities of generating directed links.
From Fig. \ref{fig45}a the strong convergence of the dynamics
of dynamic random networks towards the asymptotic dynamics
$\Phi_g^{(\infty)}$ of a strongly connected graph $g$ is apparent.
As $\Phi$ can be viewed as a random unitary quantum operation
$\Phi_{C}$ on a complete graph $C$
this is valid even for small network sizes.
However, as apparent from Fig. \ref{fig45}b
the dynamical convergence properties in static random networks
are considerably different.
This may be traced back to the fact that in these networks
convergence is expected for strongly connected graphs only
and that, according to our main result on random directed graphs,
strong connectedness arises as a
typical property only for sufficiently large graphs.

\section{Conclusion}
\label{chapter_4}
In summary, we have presented a recursive procedure for evaluating the probability
that a random directed graph is strongly connected. It reveals
that for sufficiently small numbers
of vertices this probability
is small. In this mesoscopic regime the asymptotic dynamics of qubit networks coupled by random unitary operations, for example,
	depends strongly on details of the graph characterizing its possible random couplings.
However, large random networks with constant
link-generating probability are almost certainly strongly connected.
This typical topological property has important dynamical consequences for quantum networks governed by random unitary couplings, such as
entanglement generating controlled-not operations. Irrespective whether their randomness is
static or dynamic
the probabilistic dominance of strongly connected graphs results in a characteristic
long time dynamics. It is universal in the sense that all strongly
connected qubit networks with the same number of nodes share the same long time dynamics independent of all specific differences beyond their strong connectedness. Provided all links of a large network are established with the same probability, the detailed knowledge of the actual network topology is irrelevant and thus consequently redundant. Our lack of such an information is turned into our favor and the network possess a unique asymptotic evolution.
It may be conjectured that similar typical topological properties
may also be crucial for capturing the asymptotic
dynamics of large quantum networks with dynamics different from the controlled-not
operations considered here. Thus, it is conceivable that even highly complex quantum networks,
such as a future quantum internet, could exhibit analogous universal
topological and dynamical properties.

\section*{Acknowledgments}  JN and IJ received financial support from grants No. RVO 68407700 and GACR 13-33906 S. GA acknowledges financial support by the DFG within the SFB 1119 CROSSING and by the BMBF within the project Q.com.

\end{document}